\documentclass[prl,longbibliography,aps,reprint,showpacs,raggedbottom,nofootinbib,superscriptaddress,floatfix]{revtex4-2}

\usepackage{graphicx}
\usepackage{dcolumn}
\usepackage{bm}
\usepackage{hyperref}
\usepackage{gensymb}
\usepackage{siunitx}
\usepackage{color}

\begin{document}

\title{Large-scale characterization of Cu$_2$O monocrystals via Rydberg excitons}

\author{Kerwan Morin}
\affiliation{Universit\'e de Toulouse, INSA-CNRS-UPS, LPCNO, 135 Av. Rangueil, 31077 Toulouse, France}
\author{Delphine Lagarde}
\affiliation{Universit\'e de Toulouse, INSA-CNRS-UPS, LPCNO, 135 Av. Rangueil, 31077 Toulouse, France}
\author{Angélique Gillet}
\affiliation{Universit\'e de Toulouse, INSA-CNRS-UPS, LPCNO, 135 Av. Rangueil, 31077 Toulouse, France}
\author{Xavier Marie}
\affiliation{Universit\'e de Toulouse, INSA-CNRS-UPS, LPCNO, 135 Av. Rangueil, 31077 Toulouse, France}
\author{Thomas Boulier}
\email{boulier@insa-toulouse.fr}
\affiliation{Universit\'e de Toulouse, INSA-CNRS-UPS, LPCNO, 135 Av. Rangueil, 31077 Toulouse, France}

\begin{abstract}
Rydberg states of excitons can reach microns in size and require extremely pure crystals. We introduce an experimental method for the rapid and spatially-resolved characterization of Rydberg excitons in copper oxide (Cu$_2$O) with sub-micron resolution over large zones. Our approach involves illuminating and imaging the entire sample on a camera to realize a spatially-resolved version of resonant absorption spectroscopy, without any mobile part. This yields spatial maps of Rydberg exciton properties, including their energy, linewidth and peak absorption, providing a comprehensive quality assessment of the entire sample in a single shot. Furthermore, by imaging the sample photoluminescence over the same zone, we establish a strong relationship between the spectral quality map and the photoluminescence map of charged oxygen vacancies. This results in an independent, luminescence-based quality map that closely matches the results obtained through resonant spectroscopy. Our findings reveal that Rydberg excitons in natural Cu$_2$O crystals are predominantly influenced by optically-active charged oxygen vacancies, which can be easily mapped. Together, these two complementary methods provide valuable insights into Cu$_2$O crystal properties. 
\end{abstract}

\maketitle

\section{Introduction}

Cuprous oxide (Cu$_2$O) has emerged as a fascinating material system for the study of Rydberg excitons~\cite{kazimierczuk2014giant,assmann2020semiconductor} due to its unique electronic structure and potential applications in optoelectronic devices~\cite{walther2018giant,khazali2017single,taylor2022simulation,ziemkiewicz2019solid,ziemkiewicz2023optical,walther2022nonclassical}. These excitons, comprising a bound electron-hole pair with high principal quantum numbers, up to $n=30$~\cite{versteegh2021giant} exhibit remarkable optical properties~\cite{assmann2016quantum,zielinska2016optical,morin2022self,gallagher2022microwave,heckotter2018rydberg} that are highly sensitive to local crystal environments~\cite{heckotter2017scaling, heckotter2018rydberg, kruger2020interaction}. Understanding and controlling the properties of Rydberg excitons in Cu$_2$O are crucial for developing Cu$_2$O-based devices~\cite{orfanakis2022rydberg,walther2018giant,khazali2017single,taylor2022simulation} and exploring fundamental phenomena in condensed matter physics~\cite{walther2018interactions, assmann2016quantum, walther2023quantum, walther2020plasma}.

Characterizing the spatial distribution of Rydberg exciton properties within a Cu$_2$O sample is of paramount importance as it provides critical insights into the material's quality and the impact of defects on its optical properties~\cite{kruger2020interaction, rogers2022high, lynch2021rydberg,bergen2023large,heckotter2023neutralization}. Conventional approaches for such characterization often involve scanning spectroscopy techniques that are costly, time-consuming and may not provide simultaneous information across a large area~\cite{rogers2022high}. In this article, we introduce a novel experimental method that overcomes these limitations by enabling rapid and spatially-resolved characterization of Rydberg exciton spectra over large zones.

We first present our method, that leverages resonant absorption spectroscopy in transmission mode~\cite{kazimierczuk2014giant}, where the entire sample is illuminated and imaged simultaneously, akin to hyperspectral imaging~\cite{chang2003hyperspectral}. By recording the spatial variations of Rydberg exciton spectra, we derive comprehensive spatial maps of their energy shift, linewidth, and peak absorption in a single probe laser scan. These maps constitute a direct quality assessment of an entire Cu$_2$O sample, with sub-micron resolution over millimeters, offering invaluable diagnosis of its suitability for Rydberg exciton-based research and technologies~\cite{orfanakis2022rydberg,walther2018giant,khazali2017single,taylor2022simulation,ziemkiewicz2019solid,ziemkiewicz2023optical,walther2022nonclassical}. Second, we establish a significant coincidence between this direct quality map and the photoluminescence map of charged oxygen vacancies in Cu$_2$O. Specifically, we introduce a luminescence-based quality map derived from the ratio of 1S exciton photoluminescence to oxygen vacancies photoluminescence. Remarkably, this luminescence-based map closely matches the results obtained through the direct resonant spectroscopy method, underscoring the influence of oxygen impurities on high principal quantum number Rydberg excitons. Finally, we briefly describe the effect of annealing a low-quality sample to modify the vacancy distribution and improve its excitonic properties.

\section{Experimental Methods}

\begin{figure}[t]
\begin{center}
  \includegraphics[width=0.9\linewidth]{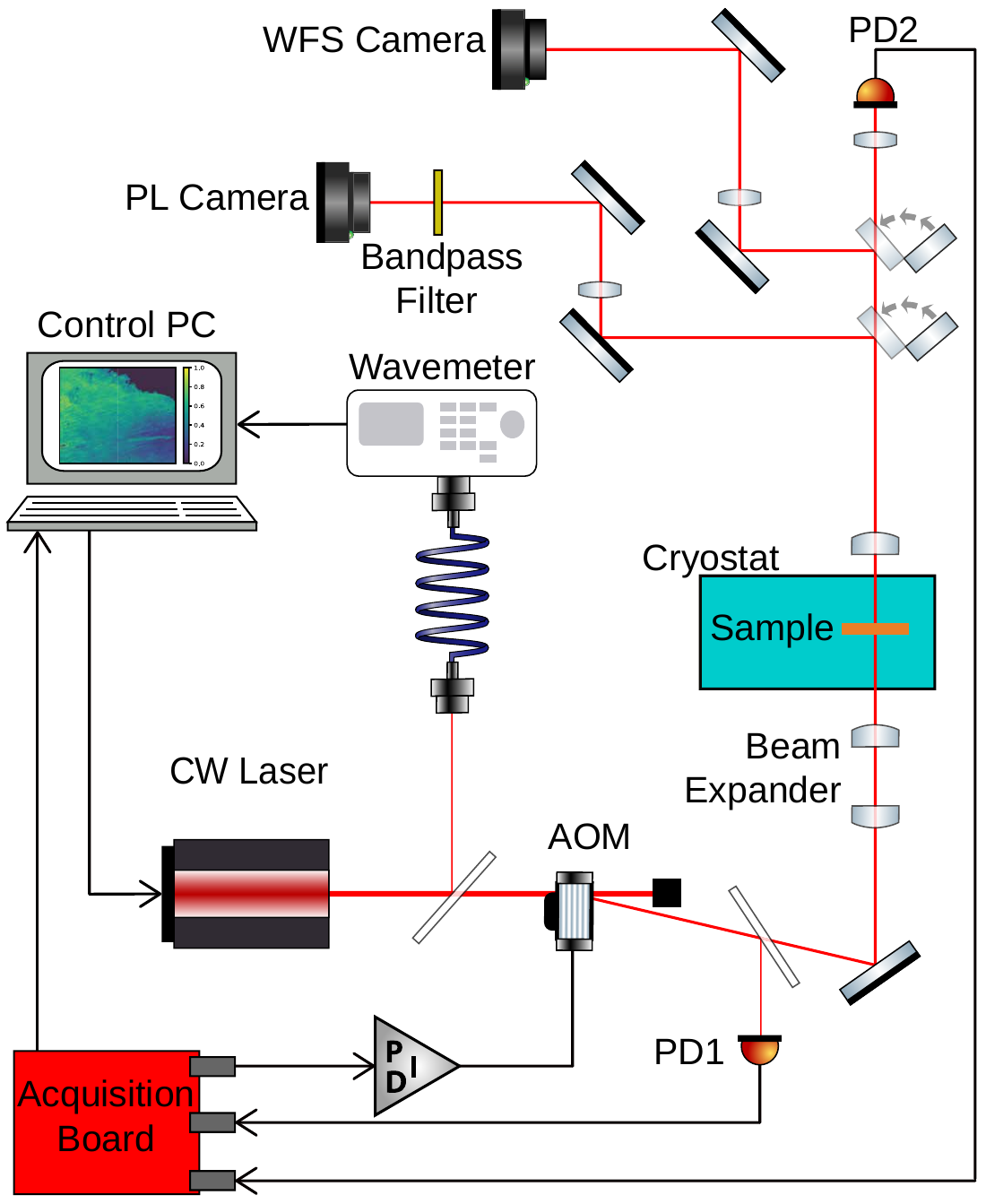}
  \caption{\textit{Schematic of the experiment}: a tunable CW laser, intensity- and frequency-locked, covers the whole sample. This sample, cooled below $3~\si{\kelvin}$, is imaged onto a camera for WFTS. The transmitted laser light can also be measured on a photodiode to get more precise transmission spectra, but at the cost of spatial resolution. The laser can be tuned to above-gap energies and the PL can be imaged as well, in which case bandpass filters select the relevant emission energies.}
  \label{fig0}
  \vspace{-0.5cm}
\end{center}
\end{figure}

\begin{figure*}[t]
\begin{center}
  \includegraphics[width=1\linewidth]{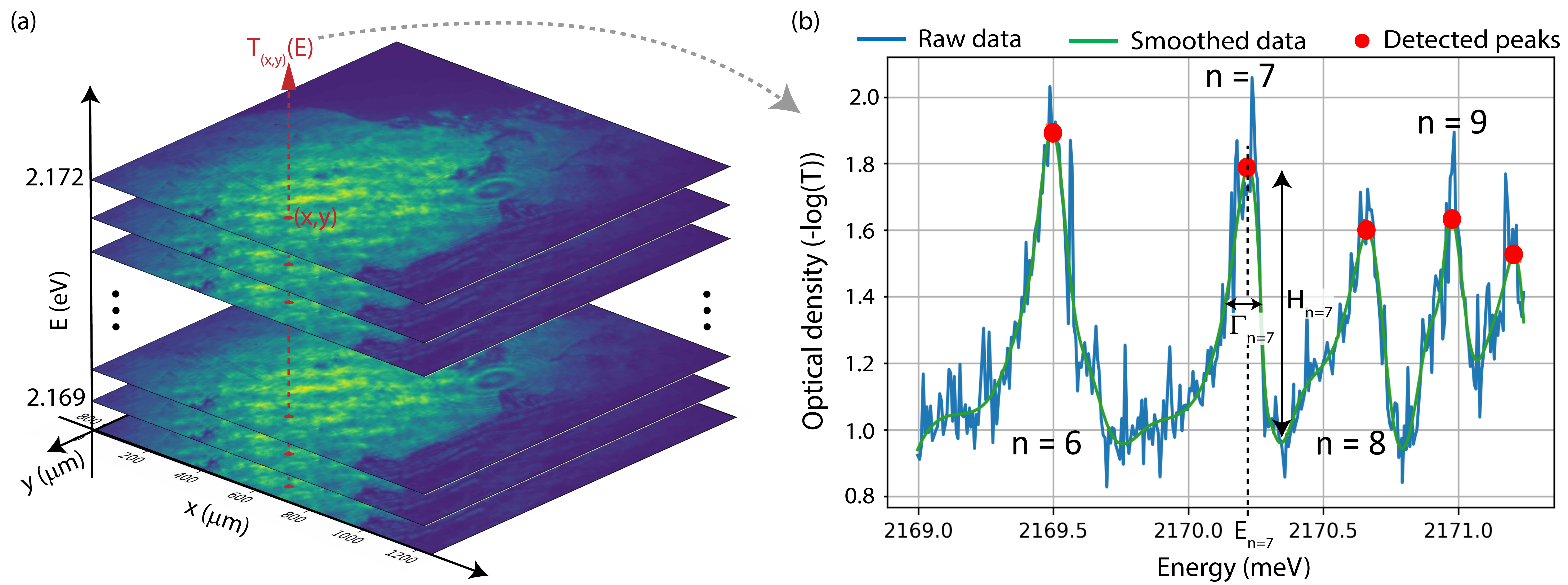}
  \caption{\textit{Graphical representation of the protocol}: (a) Imaging the sample transmission at each laser energy results in a data stack containing the Rydberg spectra for all sample positions. (b) For a given sample position, the data is carefully smoothed before each Rydberg absorption peak is detected. Their properties (FWHM $\Gamma_n$, peak height $H_n$ and peak energy $E_n$) are extracted and the process is repeated for each pixel coordinate (x,y).}
  \label{fig1}
  \vspace{-0.5cm}
\end{center}
\end{figure*}

We use commercially-procured natural Cu$_2$O crystals from the Tsumeb mine in Namibia, $50\sim 80~\si{\micro\meter}$ thick by $4\times 4~\si{\milli\meter}$ wide,  oriented such that their surface is perpendicular to the [001] crystalline axis. Both facets are highly polished so that they are optically flat. The methods described here were applied to several similar samples originating from different mother stones, so as to check the robustness of the results on crystals with various defect density and distribution. Each sample is cooled down to $2.7~\si{\kelvin}$ in a closed-cycle cryostat, where it is glued at two of its corners on a copper sample holder. In some cases, this results in a small strain gradient across the sample, which induces a local line shift but does not noticeably degrade the spectra quality~\cite{agekyan1977spectroscopic,kruger2018waveguides}.

As in represented on Figure~\ref{fig0}, narrow-linewidth ($<1~\si{\mega\hertz}$), continuous-wave (CW), doubled optical parametric oscillator (OPO) laser tunable in the $450-650~\si{\nano\meter}$ range (C-Wave model from Hubner Photonics) is used to probe the sample. It is both frequency- and intensity- locked such that frequency and power scans can be automated. The laser beam is linearly polarized and set to hit the sample at normal incidence. Throughout this paper, the laser power is adjusted such that the local intensity of the resonant light is always much below the regime of Rydberg blockade, thus avoiding optical nonlinearities. On the other side of the cryostat, the transmitted laser light is collected by a high numerical aperture (high-NA) lens and used in the detection.

In the wide-field transmission spectroscopy (WFTS) configuration, the collimated laser beam is not focused on the sample, resulting in a Gaussian-shaped illumination area of diameter $\sim 1~\si{\milli\meter}$. The transmitted light is used to image the sample plane on a Peltier-cooled monochromatic CMOS camera with a final spatial resolution of about $1~\si{\micro\meter}$ after pixel binning. The laser energy is then scanned across the yellow Rydberg series (typical step size: $1~\si{\micro\electronvolt}$) and a picture is recorded at each step. A single picture recorded at a laser energy below the relevant yellow Rydberg manifold (typically near $2~\si{\electronvolt}$, where transmission is high) is used as a reference image containing the laser spatial distribution along with the position of surface irregularities (glue drops, dust motes, scratches, ...). These irregularities are used as machine vision markers to numerically align all scan images onto the reference image, thus compensating for any potential drifts during the acquisition (or between two separate acquisition campaigns). All pictures are recorded with the same camera settings (gain, exposure time) such that the pixel values are directly comparable from picture to picture. Therefore, by dividing each scan picture by the reference picture, we recover a series of 2D arrays containing the local sample transmission at each pixel coordinate (x,y) and laser energy E, $T_{(x,y)}(E)$. This is graphically represented in Fig.\ref{fig1}(a). For a given coordinate (x,y), we can therefore recover the optical density spectrum $-log_{10}(T_{(x,y)}(E))$. As visible in Fig.\ref{fig1}(b), the pixel data from the CMOS camera is much noisier than a photodiode signal: we used an interpolating spline to smooth the data, on which a peak detection algorithm is used to extract the position, the width, and the height for each Rydberg state. We thereby recover these quantities for all detected Rydberg states, for each sample position simultaneously. From this information we build 2D maps of the full width at half maximum (FWHM) $\Gamma_n$, the absorption height $H_n$ and the energy $E_n$ for each principal quantum number $n$.

\begin{figure*}[t]
\begin{center}
  \includegraphics[width=1\linewidth]{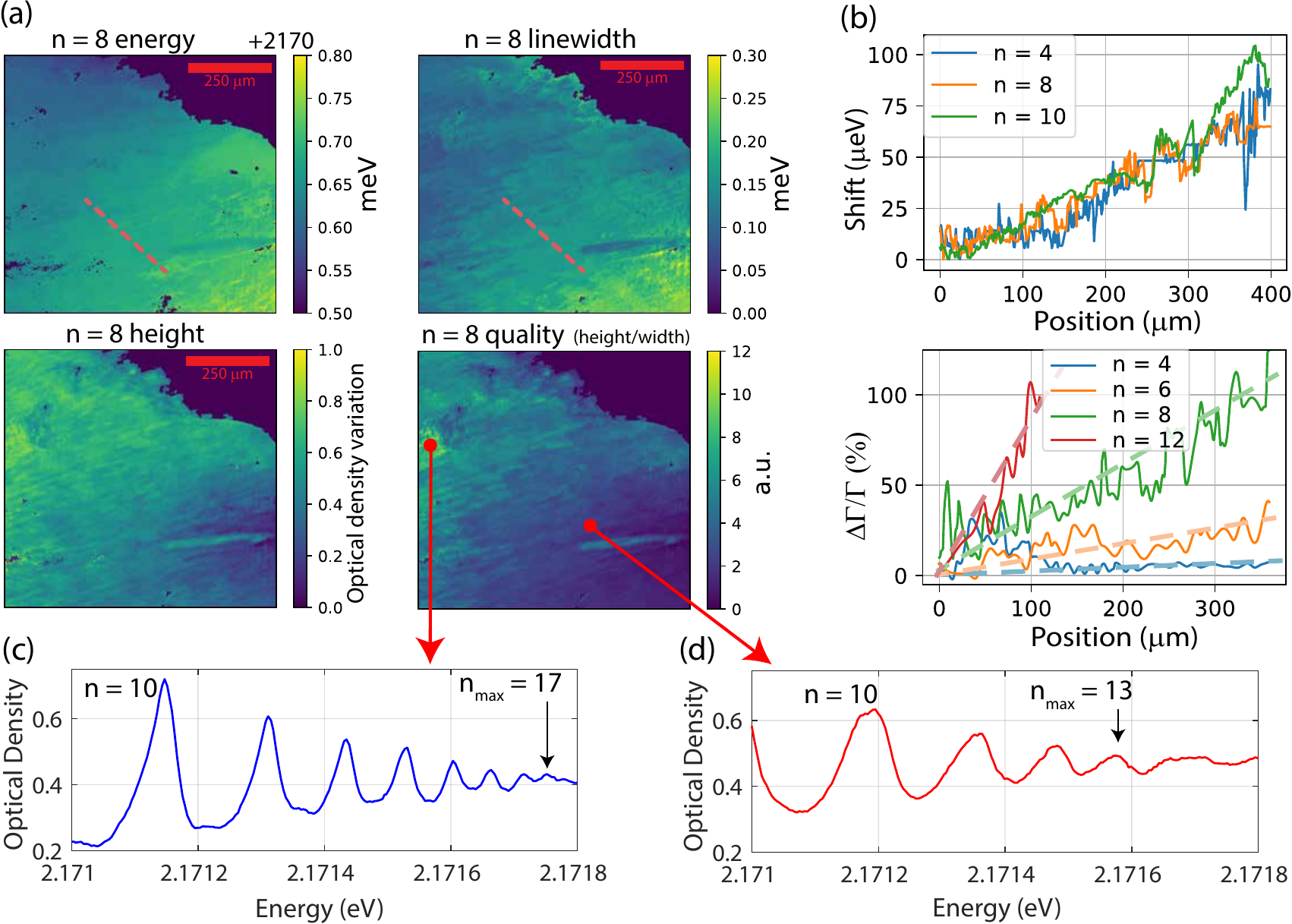}
  \caption{\textit{Exemplar maps and spectra}: (a) energy, linewidth and height WFTS maps for n = 8. The pictures diagonal spans $1~\si{\milli\meter}$. The n = 8 quality map is constructed from the latter two maps: quality = height / width. (b) Relative spatial variation of the energy (top) and linewidth (bottom, dashed lines are guides for the eye) for several principal quantum numbers n, along a cut in the gradient direction (red dashed lines in (a), top panels). The broadening $\Delta\Gamma/\Gamma$ has a strong ($n^3$) variation with n, while the energy shifts are n-independent. (c-d) Examples of spectra obtained at two sample positions, identified through the quality map as higher quality (c) and lower quality (d). Crucially, $n_{max}$ correlates well with the quality of the lower Rydberg peaks: here, from the n = 8 quality map alone we can immediately find a sample region with $n_{max}=17$, while most other sample spots only show $n_{max}=11\sim 14$.}
  \label{fig2}
  \vspace{-0.5cm}
\end{center}
\end{figure*}

We can also locally check the spectra obtained by WFTS against a more traditional, high-resolution resonant absorption spectroscopy setup, by focusing the laser on the sample (waist $\sim 10~\si{\micro\meter}$) and recording the transmitted light on a photodiode. Compared with WFTS, this reduces the detection noise at the cost of spatial resolution. We found that the WFTS spectra are systematically trustworthy up to $n=10\sim 12$ (the exact number is sample-dependant), which we found to be largely sufficient for the purpose of sample quality diagnosis.

We also looked for common features between our WFTS maps and the spatial distribution of the photoluminescence (PL). To this aim, we use a photoluminescence imaging (PLI) configuration, similar to the WFTS configuration: the laser is not focused on the sample so as to illuminate a large zone and the PL is collected in transmission mode to be imaged onto the camera. The laser is set to a wavelength of $545~\si{\nano\meter}$ ($2.275~\si{\electronvolt}$) and a long-pass filter (cutoff wavelength $575~\si{\nano\meter}$) is placed in the imaging path to block any transmitted laser light. Additional band-edge filters can be inserted before the camera to image specific PL energies. To identify the relevant energies to be imaged, the PL spectrum can be resolved using a spectrometer instead of the camera.

\begin{figure*}[t]
\begin{center}
  \includegraphics[width=0.87\linewidth]{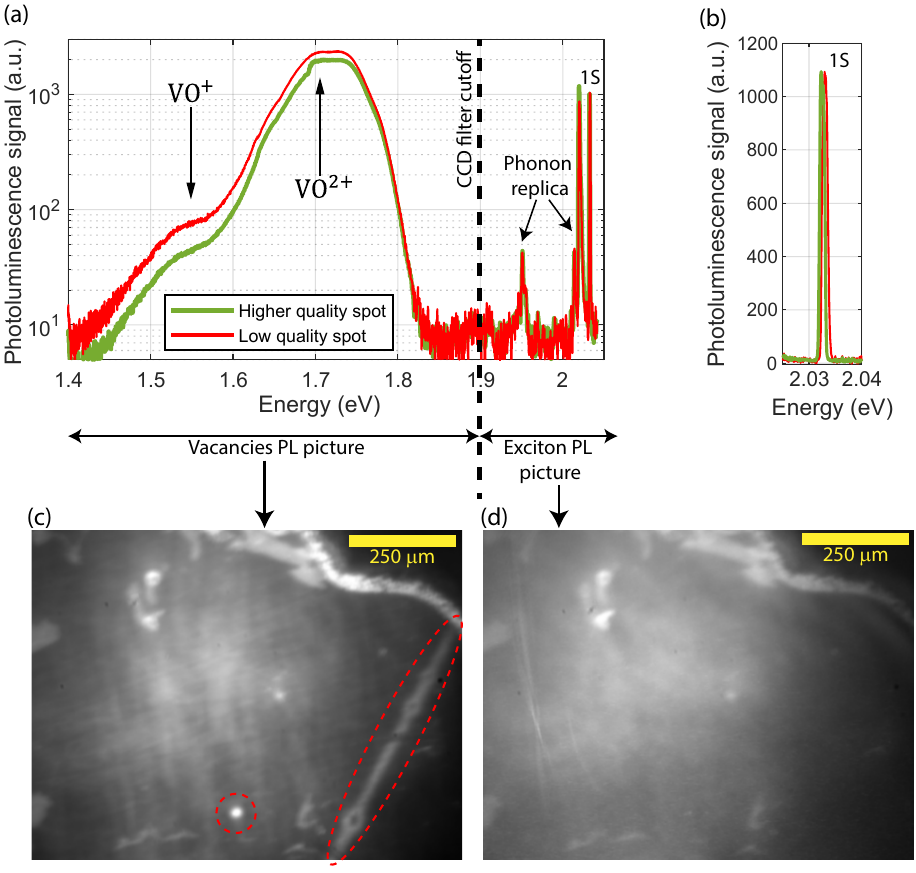}
  \caption{\textit{Photoluminescence spectra and imaging} - (a) Examples of PL spectra emitted from two sample regions of different quality. In both, a clear emission from the 1S exciton and associated phonon replica are present around $2~\si{\electronvolt}$, along with a large emission in the $1.5-1.7~\si{\electronvolt}$ region originating from oxygen vacancies (VO). The spectra are normalized to have the same 1S peak height. The lower quality region emits relatively more oxygen vacancy PL, indicating a higher density of charged defects. (b) Zoom on the 1S exciton PL. (c) Photoluminescence pictures across the sample for two spectral regions: the oxygen vacancies (left) and the 1S exciton plus phonon replica (right). The two types of emission reveal different features, such as the fault line on the right and the bright spot on the bottom (red dashed circles) that only appear in the vacancies photoluminescence.}
  \label{fig3}
  \vspace{-0.5cm}
\end{center}
\end{figure*}

\begin{figure*}[t]
\begin{center}
  \includegraphics[width=0.9\linewidth]{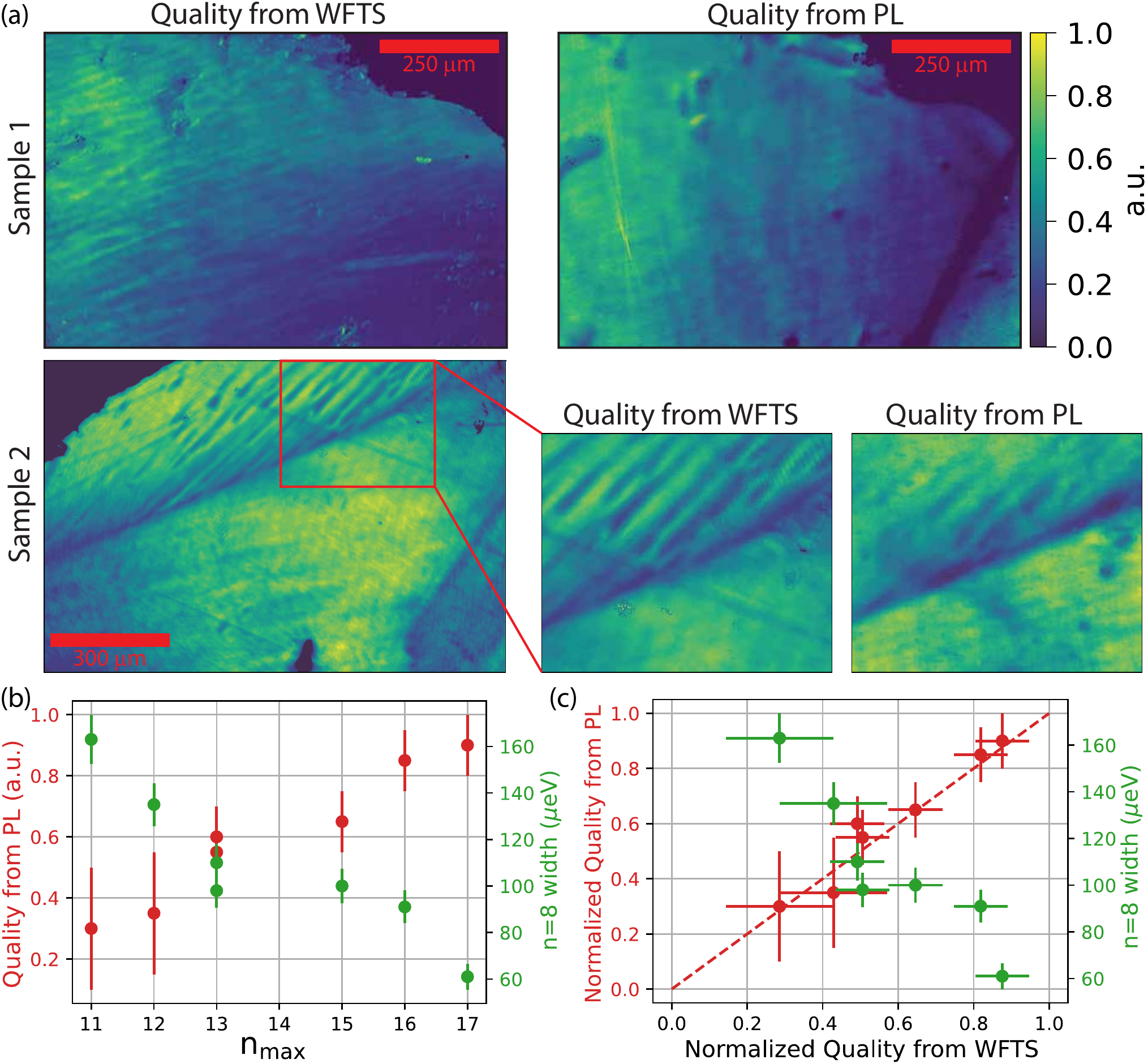}
  \caption{\textit{PL vs WFTS vs $n_{max}$} - (a) Comparison between the quality maps from direct WFTS and from PL imaging, for two different samples. Most features, such as the quality gradient in sample 1 (top), are recognizable in both. Sample 2 (bottom) shows non-trivial crystallographic features visible with both methods (zoomed-in panels). (b) Evolution of the quality estimated from PL (red) and of the $\mathrm{n=8}$ linewidth with $\mathrm{n_{max}}$ (green). (c) Visual comparison between the two quality estimates (red) where the dashed line is the unit slope, and between the quality from WFTS and the linewidth of the $\mathrm{n=8}$ peak (green).}
  \label{fig4}
  \vspace{-0.5cm}
\end{center}
\end{figure*}

\section{Results and Discussion}

\textbf{Wide field spectroscopy} - Exemplar WFTS maps (peak energy, linewidth and height) are presented in fig.~\ref{fig2} (a) for $n=8$ for a crystal (Sample 1) of medium-low quality. We can see that the linewidth and the peak height are anti-correlated, as is typical of inhomogeneous broadening. Therefore, we chose to construct a figure of merit akin to the local sample quality by computing the ratio $Q_n = H_n/\Gamma_n$, as taller and narrower absorption peaks correspond to a reduced broadening and therefore a better local crystal quality~\cite{kruger2020interaction}. As there exists a quality map for each n-state, a general quality map is built from the average of all the n-states measurable across the studied region: $Q=\overline{Q_n}$ where the average is made across n for each pixel and goes from $n=4$ to $\min_{\forall (x,y)}(n_{max}^{auto})$, the absolute minimum across the sample of the local highest state $n_{max}^{auto}$ automatically detected on the camera. The general quality map is able to precisely reveal the best sample region in a single laser scan with sub-micron resolution (limited by the imaging system resolution). We show for comparison two spectra (fig.~\ref{fig2} (c-d)) from the best region (c) and a significantly lower quality region (d) on Sample 1. In addition to displaying sharper features for all Rydberg states, the highest quality zone also shows a higher maximum principal quantum number $n_{max}$ (as verified with high-precision spectroscopy on the photodiode). In fact, we systematically observe a clear relationship between the quality estimated from relatively low states ($n<10$) and $n_{max}$. This strongly suggests that whatever causes the broadening is also responsible for the disappearance of high Rydberg states. This also highlights the importance of mapping a few low Rydberg states as a quick way to identify regions where high n-states exist, which can then be exploited for their giant nonlinearities~\cite{walther2018giant,walther2018interactions}.

Interestingly, a clear gradient is visible in all the maps presented in fig.~\ref{fig2} (a). The situation is typically more complex in other samples, an example of which is presented later. The energy gradient is attributed to a strain gradient: as it is quite linear across the map, any broadening due to energy shifts within a finite area would be constant. This is unlike what we observe in the linewidth map, where the local linewidth varies spatially by a factor up to three across the map. Moreover, the energy gradient integrated over a pixel gives at most a local shift-induced broadening of $0.3~\si{\micro\electronvolt}$, much less that the $\sim 200~\si{\micro\electronvolt}$ broadening present at some positions. Finally, we observe the same energy shift for all values of n, which is typical of strain-induced shifts. Therefore, we are confident the energy gradient is not the cause for the quality gradient. 

Further insight can be obtained from the variation the extracted parameters with n. While the energy shifts are independent from n (fig.~\ref{fig2} (b), top), we found that the relative linewidth broadening $\Delta\Gamma_n/\Gamma_n$ is strongly n-dependent (fig.~\ref{fig2} (b), bottom) and scales as $n^3$ everywhere on the sample. This scaling, along with the vanishing of the high n-states in low quality regions, is exactly what is expected from an electric field~\cite{heckotter2017scaling, heckotter2018dissociation,kruger2020interaction}. Therefore, variations in the local density of charged defects is likely the cause for the spatial variations observed in the WFTS maps, as previously suspected~\cite{kruger2020interaction, lynch2021rydberg,bergen2023large,heckotter2023neutralization}. Based on this insight, we can estimate from~\cite{kruger2020interaction} that in our sample the density of charged defects ranges from $\sim 2.10^{10}~\si{\centi\meter^{-3}}$ for the best region ($n_{max}=17$) to $\gtrsim 2.10^{11}~\si{\centi\meter^{-3}}$ for the worst regions (where $n_{max}\lesssim 12$). This matches the typical defect density found in medium-quality natural Cu2O crystals, while the highest-purity crystals have a defect density of order $10^{9}~\si{\centi\meter^{-3}}$~\cite{heckotter2020experimental}.\\

\textbf{Photoluminescence imaging} - As copper oxide has known charged defects that are optically active, we also investigated the PL of our samples. Figure~\ref{fig3} (a-b) shows the typical PL spectrum, where the well-know 1S exciton and its phonon-assisted replica are visible in the $1.9-2.05~\si{\electronvolt}$ region. A large fluorescence is also visible below $1.85~\si{\electronvolt}$, that we identify from the literature~\cite{ito1997detailed,koirala2013correlated,koirala2014relaxation,frazer2015evaluation,frazer2017vacancy,lynch2021rydberg} as charged oxygen vacancies ($\mathrm{O^+}$ vacancies at $1.55~\si{\electronvolt}$ and, $\sim 50\times$ brighter, $\mathrm{O^{2+}}$ vacancies at $1.7~\si{\electronvolt}$). We confirmed their nature with a temperature study, where we saw their photoluminescence vanish near $165~\si{\kelvin}$ due to ionization~\cite{ito1997detailed,koirala2013correlated,koirala2014relaxation,lynch2021rydberg}. Copper vacancies are also known to emit near $\sim 1.4~\si{\electronvolt}$~\cite{ito1997detailed,frazer2017vacancy,lynch2021rydberg} but we found no significant signal in this region: oxygen vacancies seem to dominate, as is expected from natural samples and unlike synthetic crystals~\cite{ito1997detailed,lynch2021rydberg}.

Comparing the PL spectra of the two sample positions previously identified with the WFTS quality map (green and red lines in fig.~\ref{fig3} (a), respectively higher and lower quality spots in fig.~\ref{fig2} (c-d)) reveals that the vacancies PL is higher in the low-quality region (by typically $25\%$), for an equal 1S emission intensity (fig.~\ref{fig3} (b)). This is in line with the assumption that the two types of charged oxygen vacancies are the main quality-degrading agents for Rydberg excitons in natural samples, at least in part due to the local electric fields these charged defect generate. Interestingly, our observation confirms the hypothesis made in previous works~\cite{heckotter2018rydberg,heckotter2020experimental,kruger2020interaction,bergen2023large} that charged impurities ultimately limit the formation of high-n excitons. 

Based on these observations, we can selectively image the intensity distribution of the vacancies (the excitons) photoluminescence using a long- (short-) pass filter with cutoff energy around $1.9~\si{\electronvolt}$. This is shown in fig.~\ref{fig3} (c-d). Interestingly, different features emerge from the two spectral domains. For example, fault lines (bright line on the right of fig.~\ref{fig3} (c)) and localized inclusions (white spot on the bottom of the same figure) are quite bright with VO PL (bright line on the right of fig.~\ref{fig3} (c)), while completely absent with exciton PL. On the other hand, surface dirt and sample edges are visible as bright features in both images. Owing to these differences, sample quality estimations from the PL will inevitably be skewed by local faults and are therefore less quantitative that the WFTS approach. Crucially, these PL pictures are not by themselves indicators of the local density of oxygen vacancies, because the pump laser illumination is not homogeneous. However, this laser inhomogeneity is exactly the same in both pictures. Thus, assuming the PL intensity is linear in pump intensity, their (pixel-by-pixel) ratio is exempt from pump background and provides a map of the relative PL contributions from excitons and vacancies. Figure~\ref{fig4} (a) (top right) shows such a map for the ratio $Q_{PL}=I_{ex}/I_{VO}$ of excitonic PL over vacancies PL, for which higher values mean less defects. To check whether this correlates with higher quality regions, the general quality map obtained from WFTS is also presented in figure~\ref{fig4} (a) (top left). The agreement is good, with the quality gradient appearing in $Q_{PL}$ as well. Furthermore, we measured $n_{max}$ at several spots with the high precision (photodiode) transmission spectroscopy and compared it with the two quality estimates. The result is shown in figure~\ref{fig4} (b), which displays a high correlation between $n_{max}$ and the quality from PL, as well as between $n_{max}$ and an exemplar width (here for $n=8$). Finally, from this same data, figure~\ref{fig4} (c) shows the good agreement between the two quality estimates. Therefore, selective photoluminescence imaging is also able to rapidly pinpoint good quality regions within a mixed-quality sample. Notice, however, that a detailed comparison shows significant differences. First, as noted above, the two spectral regions making up the PL-based quality map have widely different response to faults and other large defects. This results in such features being exaggerated in the PLI map. Second, the green laser pump used for photoluminescence has a much shorter penetration length than the yellow laser used for absorption spectroscopy. Therefore, the two methods do not explore the same crystal depths: WFTS probes the whole crystal thickness while PLI mostly explores crystal depths of a few microns on the side of the laser pump. This explains why WFTS is able to resolve finer structures and is less sensitive to surface defects, as is apparent in figure~\ref{fig4} (a).

We stress that the two types of quality estimation (WFTS and PLI) are completely independent and that their agreement is therefore significant: first, it confirms that the main hindrance to high Rydberg states of excitons in natural crystals is specifically $\mathrm{VO^{(2)+}}$ vacancies. Second, it provides an even quicker and more cost-effective way than WFTS to identify samples suitable for solid-state Rydberg physics, a crucial step for developing Rydberg exciton-based technologies in the current absence of any systematic source of high-quality sample (both synthetic and natural)~\cite{lynch2021rydberg}. 

Figure~\ref{fig4} (a)(bottom) shows an example from a higher-quality sample (Sample 2), where interesting crystal features are visible above a fault line. While their exact nature in unknown, it is noteworthy that they do not appear in the exciton part of the PL, nor in white light (transmission, reflection), but are visible in both quality maps. This means they are not localized voids nor inclusions, which would show up in white light, but a more subtle crystallographic property that correlates with local VO doping. As was the case for Sample 1, while both WFTS and PLI approaches provide qualitatively similar information, the PLI quality map is typically less precise and explores a limited crystal depth.\\

\begin{figure}[t]
\begin{center}
  \includegraphics[width=0.9\linewidth]{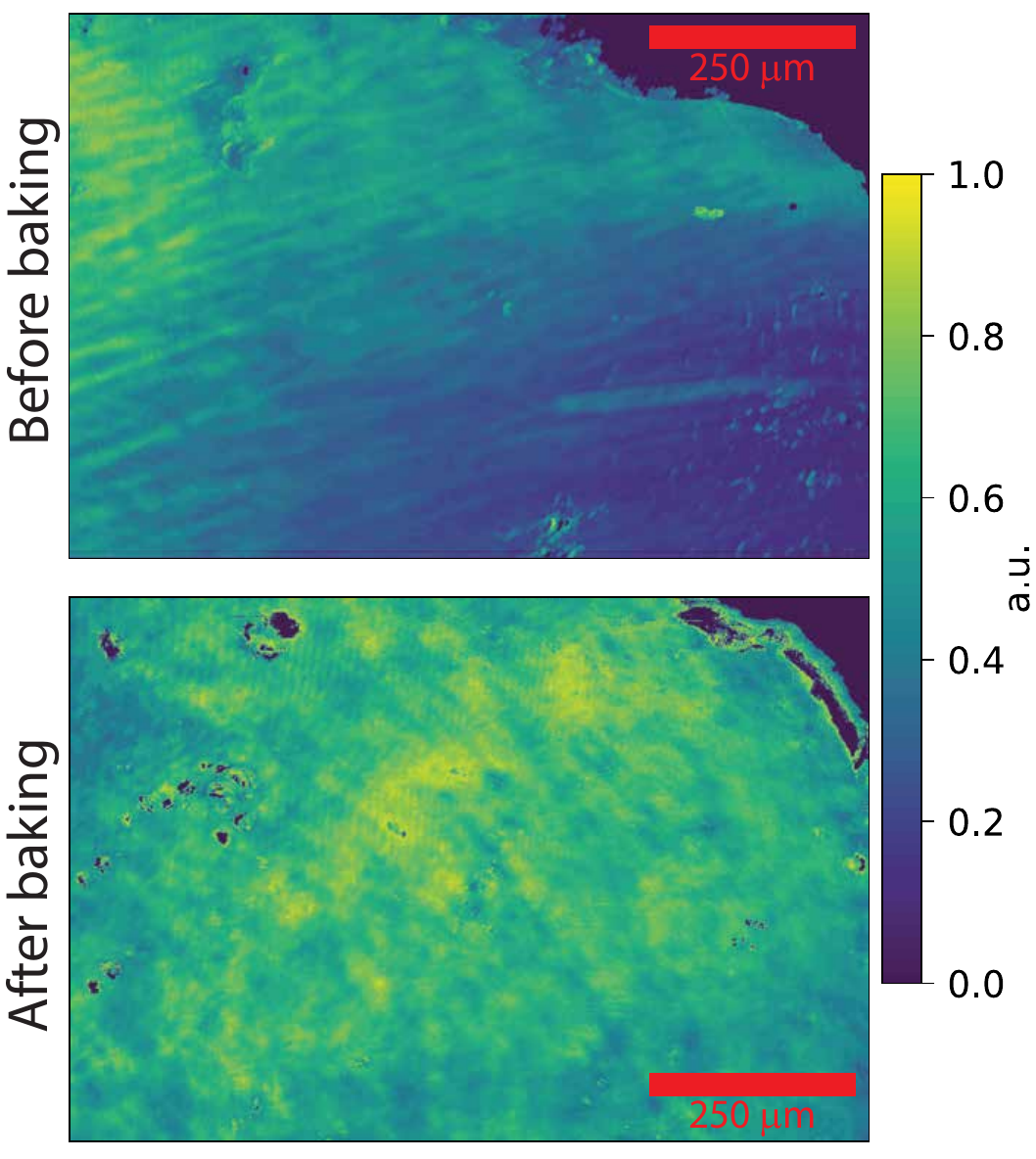}
  \caption{Quality map of the same sample (Sample 1) before (top) and after (bottom) the annealing process. The quality gradient, which we attribute to a density gradient of oxygen vacancies, disappeared and the average quality is thereby increased on large scales. Local variations remain, and we only observe a marginal improvement on $n_{max}$.}
  \label{fig5}
  \vspace{-0.5cm}
\end{center}
\end{figure}

\textbf{Annealing} - Finally, we attempted to improve the quality of Sample 1 with an annealing process, designed to modify the density and the distribution of oxygen defects. The procedure consists in baking the natural sample at $290~\si{\celsius}$ for five days in a $100\%$ $O_2$ atmosphere at ambient pressure. This slow annealing at low temperature avoids oxidation into CuO at ambient pressure~\cite{wang2014transmittance}. Meanwhile, the oxygenated atmosphere increases the oxygen-to-copper ratio and avoids copper vacancies~\cite{chang2013removal}, in an attempt to remove oxygen vacancies while keeping copper vacancies low. Figure~\ref{fig5} compares the general quality map of Sample 1 obtained with WFTS before and after the procedure. The linear gradient has disappeared, the quality is globally improved with larger zones of relatively good quality, while large-scale variations are still present. However, we note that this baking procedure damaged the surface, with an average roughness jumping from $<10~\si{\nano\meter}$ (before) to $\lesssim 1~\si{\micro\meter}$ (after), as measured by white light interferometry. This implies that some of the smallest features in the after-bake map may be artefacts from the surface roughness. Nonetheless, this demonstrates that WFTS is still able to produce quantitative results at least on larger scales, despite the sub-optimal sample condition. Annealing natural crystals of medium-low quality in this fashion may become a viable solution to the current problem of high-purity crystal sourcing, provided the surface damage can be avoided. Tentative solutions, beyond the scope of this paper, include using a lower temperature, reducing the $O_2$ concentration or, more radically, re-polishing the sample after baking.

\section{Conclusion}

In conclusion, we have presented a novel experimental method for the rapid and spatially-resolved characterization of Rydberg exciton spectra in copper oxide (Cu$_2$O). Our approach, based on resonant absorption spectroscopy and photoluminescence mapping, offers unique advantages for assessing the quality and properties of Cu$_2$O crystals. 
The spatially-resolved resonant absorption spectroscopy method allows to obtain comprehensive maps of Rydberg exciton properties, including their energy shift, linewidth, and peak absorption, all in a single laser energy scan. This technique provides invaluable insights into the overall optical quality of Cu$_2$O samples and can reveal local variations due to defects or impurities. Furthermore, the correlation between the resonant absorption-based quality map and the photoluminescence-based quality map highlights the significant influence of charged oxygen vacancies on high principal quantum number Rydberg excitons in natural Cu$_2$O crystals. This finding not only helps identify zones of high quality within a sample but also suggests avenues for improving crystal to minimize impurities, such as annealing natural crystals.



In summary, our study not only introduces an innovative experimental approach for spatially-resolved Rydberg exciton characterization but also paves the way for future investigations into the intricate interplay between crystal quality, defects, and Rydberg exciton dynamics.\\

\begin{acknowledgments}
The authors are thankful to Lise-Marie Lacroix and Pierre Renucci for useful discussions. This work has been supported through the ANR grant ANR-21-CE47-0008 (PIONEEReX), through the EUR grant NanoX ANR-17-EURE-0009 in the framework of the "Programme des Investissements d’Avenir" and through T. Boulier's Junior Professor Chair grant ANR-22-CPJ2-0092-01.
\end{acknowledgments}

\bibliographystyle{apsrev4-2}
\bibliography{biblio}

\end{document}